# Crystal structure, cobalt and iron speciation and oxygen non-stoichiometry of $La_{0.6}Sr_{0.4}Co_{1-y}Fe_yO_{3-\delta}$ nanorods for IT-SOFC cathodes

*Augusto E. Mejía Gómez [a], Joaquín Sacanell [b,c], Cristián Huck-Iriart [d], Cinthia Ramos [b,c], Analía L. Soldati [c], Santiago J. A. Figueroa [e], Manfredo H. Tabacniks [f], Márcia C. A. Fantini [f], Aldo F. Craievich [f] and Diego G. Lamas [d,\*]*

[a] Grupo de Películas Delgadas y Nanofotónica, Departamento de Física, Pontificia Universidad Javeriana, Bogotá, Colombia

[b] Departamento Física de la Materia Condensada, Centro Atómico Constituyentes, Comisión Nacional de Energía Atómica, San Martín, Pcia. de Buenos Aires, Argentina

[c] Instituto de Nanociencia y Nanotecnología (INN), CNEA-CONICET, Argentina

[d] CONICET / Laboratorio de Cristalografía Aplicada, Escuela de Ciencia y Tecnología, Universidad Nacional de Gral. San Martín, San Martín, Pcia. de Buenos Aires, Argentina

[e] CNPEM, Laboratório Nacional de Luz Síncrotron, Campinas, SP, Brazil

[f] Instituto de Física, Universidade de São Paulo, São Paulo, SP, Brazil

**Abstract**

Single-phased $La_{0.6}Sr_{0.4}Co_{1-y}Fe_yO_{3-\delta}$ (y = 0.2, 0.5, 0.8) nanorods exhibiting the rhombohedral perovskite-type phase were synthesized by a pore-wetting technique. We studied their chemical composition, crystal and electronic structures, morphology and hyperfine properties as a function of the Co/Fe content of the samples. Our results demonstrate that Co cations exhibit a slightly lower oxidation state than Fe ones, resulting in a higher oxygen non-stoichiometry $\delta$ for Co-rich samples. In addition, the values of $\delta$ determined in this work for nanostructured samples are much higher than those reported in the literature for bulk materials. This can be attributed to the high degree of defects in nanomaterials and is probably one important factor in the high electrochemical performance for the oxygen reduction reaction of nanostructured $La_{0.6}Sr_{0.4}Co_{1-y}Fe_yO_{3-\delta}$ IT-SOFC cathodes, which have been reported in a previous work.

*Keywords:* electrode materials; nanostructured materials; X-ray diffraction; NEXAFS; Mössbauer spectroscopy

[\*] Corresponding author: Escuela de Ciencia y Tecnología, Universidad Nacional de General San Martín, Martín de Irigoyen 3100, Edificio Tornavía, Campus Miguelete, (1650) San Martín, Pcia. de Buenos Aires, Argentina. E-mail address: diegoglamas@gmail.com (Diego G. Lamas)

# 1. Introduction

Fuel cells are one of the most promising devices for environmentally clean energy production by directly converting chemical energy into electricity. Among them, solid-oxide fuel cells (SOFCs) have the unique capability to use different fuels such as hydrocarbons or hydrogen. However, several issues have to be solved in order to improve their efficiency and reduce their costs. The reduction of their operating temperature, which is typically around 900-1000°C, is one of the most important concerns. For this reason, extensive research has been devoted to develop novel materials for intermediate temperature SOFCs (IT-SOFCs). In the case of the cathode, the use of mixed ionic and electronic conductors (MIECs) is usually preferred, because the oxygen reduction reaction (ORR) can take place at the entire surface of the cathode, while in conventional electronic conductors it only occurs in the triple phase boundary between oxygen, electrolyte and cathode. In particular, $La_{0.6}Sr_{0.4}Co_{1-y}Fe_yO_3$ (LSCF) perovskites have shown excellent electrocatalytic properties for the ORR.

In the last years, extensive research has been devoted to study the structural and electrochemical properties of nanostructured MIECs as cathodes for IT-SOFCs [1,2,3,4,5,6,7,8]. These materials are very interesting because the number of active sites for the ORR is expected to increase dramatically due to the increase of the specific surface area. Accordingly, in a recent work, we investigated the properties of LSCF cathodes prepared from tubes and rods consisting of softly-agglomerated nanoparticles [4]. These architectures for SOFC cathodes exhibited much higher performance than conventional microstructured materials and nanorods were found to have better electrochemical properties than nanotubes. This enhanced performance was mainly attributed to the fact that these structures provide electrodes with high surface to volume ratio, increasing the number of reaction sites with the surrounding gas, when compared to an ordinary microstructured cathode. In addition, we demonstrated that the diffusivity of oxide ions increases for decreasing grain size. It is worth to point out that nanomaterials are not employed in conventional SOFCs since grain growth is expected to occur at the high operation temperatures of these devices. However, their use in IT-SOFCs has been evaluated in the last years, since these devices operate at lower temperatures (typically in the 500-700°C range).

Lowering the characteristic dimensions of materials down to nanometric level increases their structural disorder, which usually enhances electrochemical response but, however, at the same time other relevant physico-chemical properties may be seriously affected. For this reason, besides the electrochemical study carried out in our previous work, detailed investigations on their crystal, microstructural and electronic structures are mandatory. For this reason, the present work aims at characterizing the relevant properties of a LSCF nanorods of varying Co/Fe relative content. The characterization was performed by Rutherford Backscattering Spectrometry (RBS), Synchrotron X-ray Powder Diffraction (SXPD), Scanning Electron Microscopy (SEM), X-ray Absorption Near Edge Structure (XANES) spectroscopy and Mössbauer spectroscopy.

## 2. Experimental Procedure

### 2.1. Synthesis of LSCF nanorods

LSCF nanorods were synthesized by a pore-wetting technique starting from a nitrate solution of the desired cations and using polycarbonate porous membranes as templates [9,10,11].

A 1 M stoichiometric solution of $La(NO_3)_3 \cdot 6H_2O$, $Sr(NO_3)_2$, $Co(NO_3)_2 \cdot 6H_2O$ and $Fe(NO_3)_3 \cdot 9H_2O$ was prepared by the dissolution of analytical reagents in pure water. Templates of porous polycarbonate films (Millipore) were used as filters in an adequate system for syringe filtration. By this process the total volume of the pores is filled with the solution. Filters with average pore size of $\phi = 200$ nm were used in order to yield nanorods, since previous studies in other materials showed that nanotubes are obtained if filters of larger pore sizes are used. The reaction to obtain the desired compound proceeds by the denitration process of the confined precursor in a microwave oven. By adjusting the time and the energy applied to the sample it is possible to accomplish this reaction without producing damage to the polycarbonate film. Finally, the template is sacrificed during a thermal treatment in a standard furnace with a final temperature of 800°C for 10 min. An additional thermal treatment at 1000°C was required in order to obtain a single-phased material.

### 2.2. Ion Beam Analysis

The elemental composition of the samples was measured by Particle Induced X-ray Emission, PIXE, using a proton beam with 2.4 MeV, Rutherford Backscattering Spectrometry, RBS and Resonant Elastic Scattering, this last one to enhance the detection of oxygen using the $^{16}O\,(\alpha, \alpha)^{16}O$ resonant reaction at 3.038 MeV (Gurbich

A.F., NIM B371, 2016). The measurements were made in the Laboratório de Análise de Materiais por Feixes Iônicos (LAMFI-USP). PIXE quantitative calibration was done using single element thin films evaporated on Mylar, and the X-ray spectra were analyzed using *WinQxas* [12]. RBS and EBS spectra were analyzed using the *SIMNRA* code [13].

*2.3. Synchrotron X-ray Powder Diffraction (SXPD)*

SXPD experiments were carried out using the D10B-XPD beamline of the Brazilian Synchrotron Light Laboratory (LNLS, Campinas, Brazil). A high-intensity (low-resolution) configuration was used in order to detect weak Bragg peaks. The wavelength was set at 1.61017 Å.

Rietveld refinements were performed using the *FullProf* code [14], assuming that the samples exhibited the rhombohedral phase, space group $R\bar{3}c$, with ($La^{3+}$, $Sr^{2+}$) cations, ($Co^{3+}$, $Fe^{3+}$) cations and $O^{2-}$ anions in 6a, 6b and 18e positions, respectively. The peak shape was assumed to be a pseudo-Voigt function. The background of each profile was adjusted by a six-parameter polynomial function in $(2\theta)^n$, $n = 0-5$. Isotropic atomic temperature parameters were used. Given the fact that we found a strong correlation between $B(O)$ thermal parameter and the occupancy of $O^{2-}$ anions, we decided to fix the occupancy assuming the oxygen non-stoichiometry determined from XANES results (see below). The thermal parameters corresponding to La and Sr atoms (A site) were assumed to be equal, as those of Co and Fe atoms (B site). No other constraints or restraints were used. The scattering factors were corrected considering their anomalous parameters for the wavelength used in our experiments (http://skuld.bmsc.washington.edu/scatter/).

The crystallite size of all the solid solutions was determined by means of the Scherrer equation using the first intense peak at low angle to minimize the possible effect of microstrains. The instrumental broadening subtraction was performed from the analysis of a $LaB_6$ standard (NIST-SRM 660a).

*2.4. Scanning Electron Microscopy (SEM)*

The morphology of LSCF nanomaterials was examined by SEM using a FEI QUANTA 200 (Laboratorio de Microscopía Electrónica, Gerencia de Materiales, Centro Atómico Constituyentes, CNEA, Argentina) and a FEI QUANTA 250 (Centro de Investigación y Desarrollo en Mecánica, INTI, Argentina) microscopes.

*2.5. X-ray Absorption Near Edge Structure (XANES) spectroscopy*

XANES studies at the Co and Fe K-edges (7709 and 7112 eV, respectively) were carried out at the D08B-XAFS2 beamline of the LNLS (Brazil) using a Si(111) monochromator crystal. All spectra were measured in transmission mode at room temperature (RT).

For sample preparation, the powdered materials were suspended in 2-propanol and deposited on Millipore membranes. The thicknesses were adjusted to obtain a total absorption above the edge of 1.5. Co and Fe reference materials with different oxidation states were prepared and measured in the same way. Metallic foils were used as a reference in each case for energy shift (calibration/correction). A combination of Athena software and python scripting was employed for normalization and data treatment.

*2.6. Mössbauer spectroscopy*

$^{57}$Fe Mössbauer spectra were taken at room temperature in a conventional constant acceleration spectrometer in transmission geometry with a $^{57}$Co/Rh source (Departamento Física de la Materia Condensada, Gerencia Investigación y Aplicaciones, Centro Atómico Constituyentes, CNEA, Argentina). Measurements were performed at two velocity ranges (6 mm/s and 10 mm/s). The low velocity range was appropriate to analyze in detail the paramagnetic character samples. Spectra were fitted by using the *WinNormos* program [15]. Isomer shift (IS) values were given relative to that of metallic Fe at room temperature, providing information not only on the average charge state of the iron ions but also on the oxygen non-stoichiometry of the material.

## 3. Results and discussion

*3.1. Chemical analysis by RBS*

Table 1 summarizes the results of the elemental analysis performed by Ion Beam Analysis (IBA). Data have been normalized assuming the total atomic concentration in the perovskite $ABO_3$ structure is unity for both, A and B sites. Experimental results are close to the nominal concentrations.

| Nominal composition | La content | Sr content | Co content | Fe content |
|---|---|---|---|---|
| $La_{0.6}Sr_{0.4}Co_{0.8}Fe_{0.2}O_{3-\delta}$ | 0.583 (3) | 0.417 (4) | 0.797 (2) | 0.203 (2) |
| $La_{0.6}Sr_{0.4}Co_{0.5}Fe_{0.5}O_{3-\delta}$ | 0.584 (5) | 0.412 (4) | 0.487 (8) | 0.513 (6) |
| La$_{0.6}$Sr$_{0.4}$Co$_{0.2}$Fe$_{0.8}$O$_{3-\delta}$ | 0.564 (5) | 0.436 (4) | 0.189 (5) | 0.811 (6) |

**Table 1:** Results of the chemical analysis by IBA, normalized assuming that the total atomic concentration in the perovskite ABO$_3$ structure is unity for both A and B sites.

*3.2. SXPD data and Rietveld analysis*

Qualitative analysis of our SXPD data demonstrated that all LCSF materials fired at 800°C exhibited the presence of secondary phases, but pure single-phased materials can be obtained after calcination at 1000°C for one hour (see Figure 1). These samples exhibited the expected rhombohedral perovskite-type phase for all compositions. The average crystallite size, determined by the Scherrer equation, was about 25-28 nm in all cases (see Table 2).

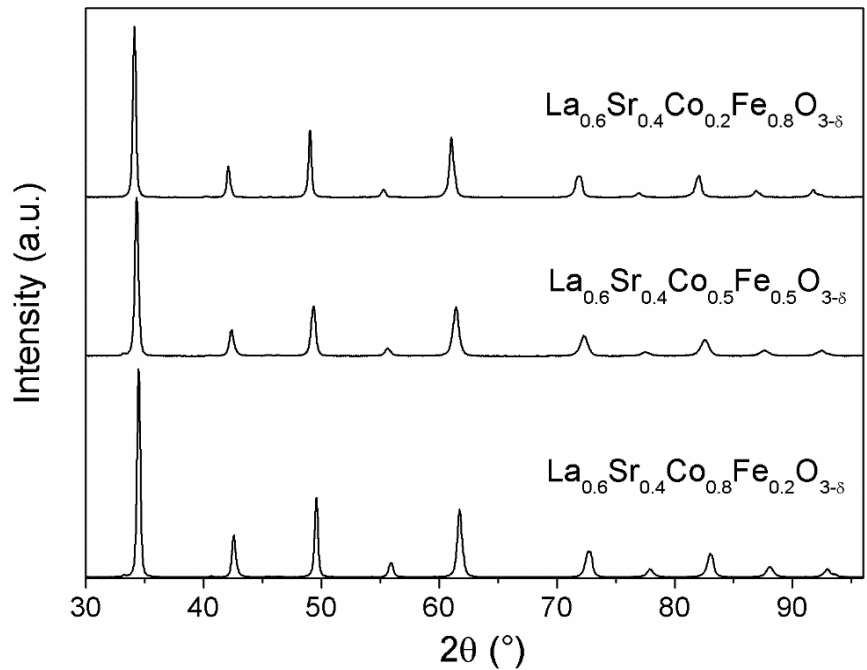

**Figure 1:** Experimental SXPD patterns corresponding to the LSCF samples calcined at 1000°C.

Figure 2 shows the SXPD pattern corresponding to the La$_{0.6}$Sr$_{0.4}$Co$_{0.2}$Fe$_{0.8}$O$_{3-\delta}$ sample and the calculated profile obtained by Rietveld analysis under the $R\bar{3}c$ space group, showing an excellent agreement among them. We observed the presence of some weak reflections, which are characteristic of the rhombohedral phase, thus confirming our hypothesis. The same was found for the other samples of the series.

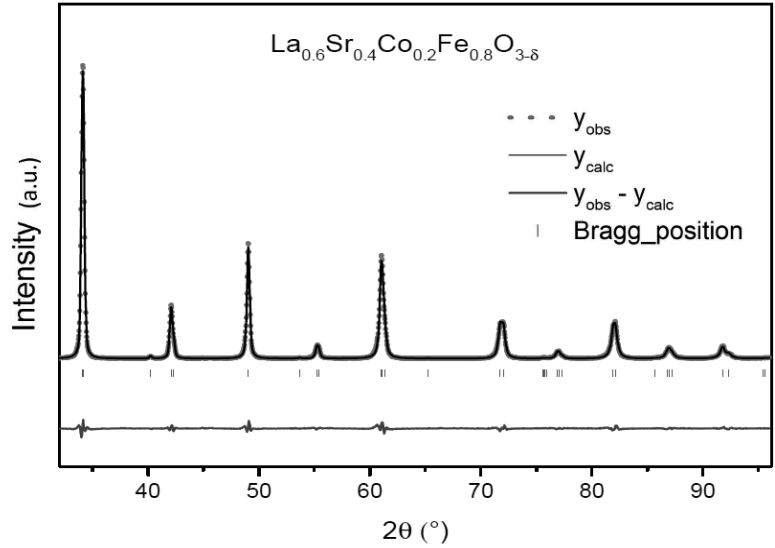

**Figure 2:** Experimental SXPD pattern and fitting of the data, obtained after Rietveld refinements for the $La_{0.6}Sr_{0.4}Co_{0.2}Fe_{0.8}O_{3-\delta}$ sample. Data, fit and difference are indicated in red, black and blue, respectively. Goodness of fit: 1.88.

The results of all the structural parameters determined by Rietveld refinements for the three compositions studied in this work are presented in Table 2. The cell parameters in hexagonal axes, *a* and *c*, increase monotonously with increasing Fe content (see Figure 3), as expected according to the values corresponding to pure ferrite and pure cobaltite compounds, $La_{0.6}Sr_{0.4}FeO_{3-\delta}$ [16] and $La_{0.6}Sr_{0.4}CoO_{3-\delta}$ [17], respectively. The fractional *x*-coordinate of the $O^{2-}$ anion in the asymmetric unit of the hexagonal unit cell, *x(O)*, and the isotropic Debye-Waller factors exhibit small changes as functions of Fe content.

|  | $La_{0.6}Sr_{0.4}Co_{0.8}Fe_{0.2}O_{3-\delta}$ | $La_{0.6}Sr_{0.4}Co_{0.5}Fe_{0.5}O_{3-\delta}$ | $La_{0.6}Sr_{0.4}Co_{0.2}Fe_{0.8}O_{3-\delta}$ |
|---|---|---|---|
| Space group | $R\bar{3}c$ | $R\bar{3}c$ | $R\bar{3}c$ |
| *a* (Å) | 5.4412 (2) | 5.4505 (5) | 5.4962 (2) |
| *c* (Å) | 13.2559 (7) | 13.321 (2) | 13.3917 (5) |
| *x(O)* | 0.530 (1) | 0.528 (2) | 0.537 (1) |
| *B(La;Sr)* (Å$^2$) | 0.33 (3) | 0.33 (4) | 0.17 (2) |
| *B(Co;Fe)* (Å$^2$) | 0.22 (7) | 0.22 (6) | 0.10 (4) |
| *B(O)* (Å$^2$) | 1.1 (1) | 1.2 (1) | 0.40 (8) |
| $\chi^2$ | 2.53 | 2.61 | 1.88 |
| *D* (nm) | 28 (3) | 25 (3) | 28 (3) |

**Table 2:** Summary of the results obtained from SXPD analysis. *a* and c are the cell parameters in hexagonal axes determined by Rietveld refinements, *x(O)* is the fractional *x*-coordinate of the $O^{2-}$ anion in the asymmetric unit of the hexagonal unit cell, *B(La;Sr) B(Co;Fe)* and *B(O)* are the isotropic Debye-Waller factors and $\chi^2$ is the goodness of fit. D is the average crystallite size determined by the Scherrer equation. See text for details.

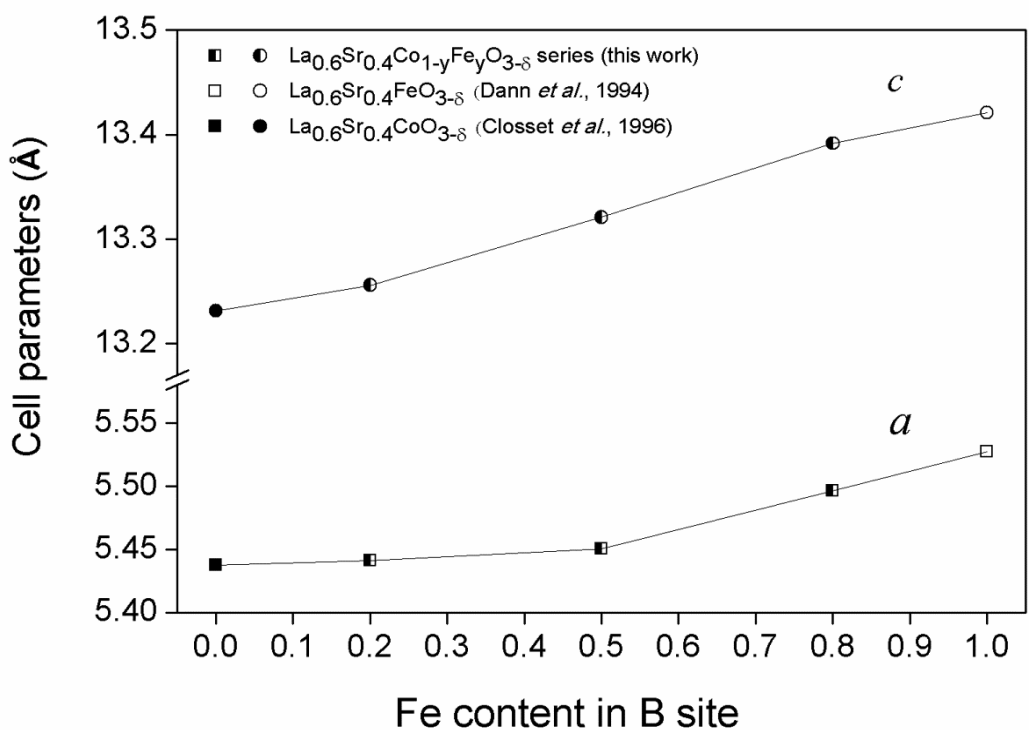

**Figure 3:** Cell parameters of the rhombohedral phase, in hexagonal axes, for LSCF nanorods. Data from the literature for $La_{0.6}Sr_{0.4}FeO_{3-\delta}$ and $La_{0.6}Sr_{0.4}CoO_{3-\delta}$ compounds were included for purpose of comparison.

*3.3. Sample morphology*

Figure 4 displays SEM micrographs of the LSCF samples after calcination at 1000ºC for 1 h, showing their rod-like nanostructure. It can be observed that similar morphologies were found in all cases, with no dependence on composition. The rods exhibited typical lengths of about 0.8-1 μm and were formed by nanoparticles with diameters in the range of 100-180 nm, with average particle size of about 140 nm independently of the sample composition. Since these nanoparticles are larger than the average crystallite sizes determined from SXPD data, it can be assumed that the particles are aggregates of crystallites. Additional discussions on the morphology of nanostructured samples synthesized by the pore-wetting technique can be found in our previous paper [4].

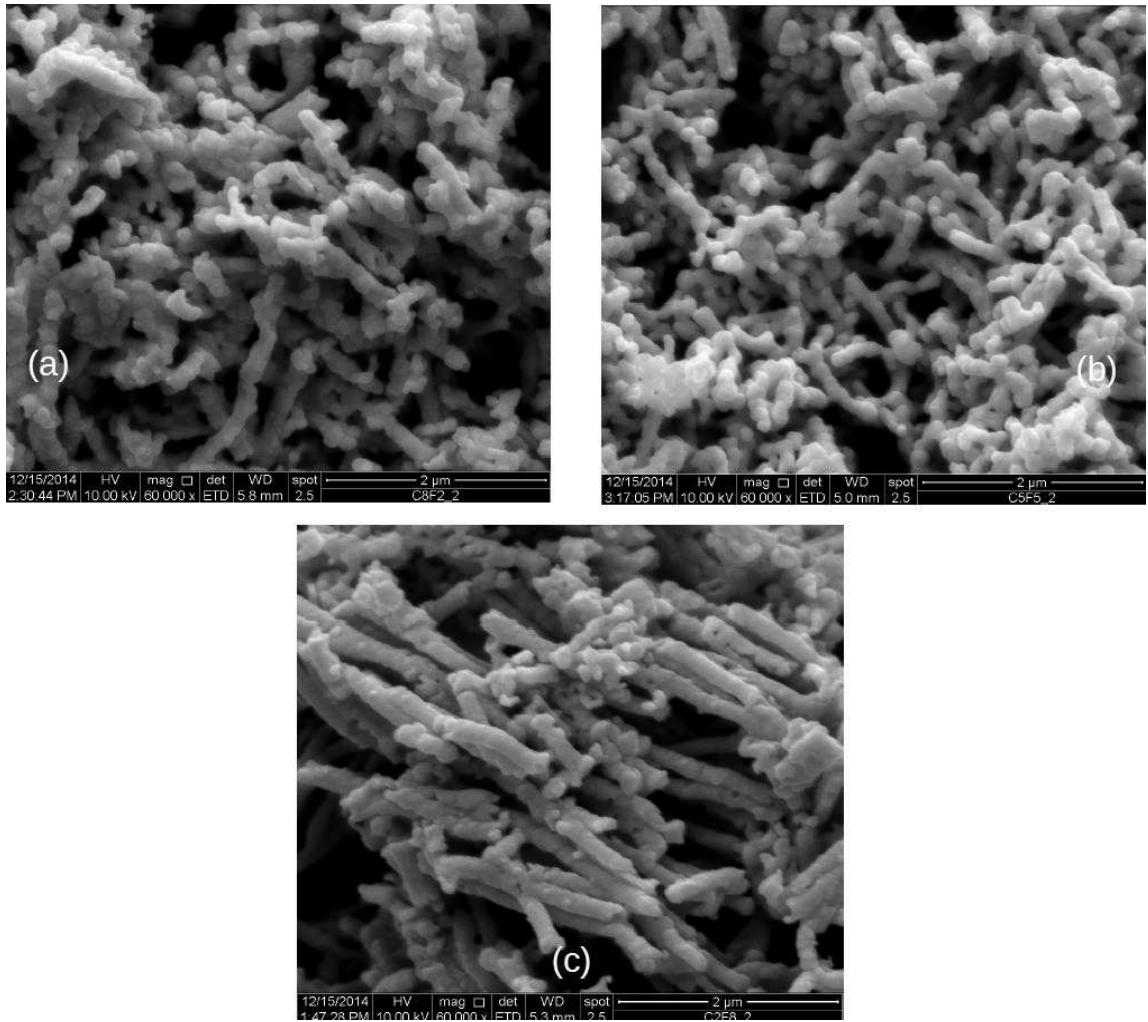

**Figure 4:** Rod-like nanostructure of LSCF samples. (a) $La_{0.6}Sr_{0.4}Co_{0.8}Fe_{0.2}O_{3-\delta}$; (b): $La_{0.6}Sr_{0.4}Co_{0.5}Fe_{0.5}O_{3-\delta}$; (c) $La_{0.6}Sr_{0.4}Co_{0.2}Fe_{0.8}O_{3-\delta}$.

*3.4. XANES analysis*

Figure 5 shows XANES spectra and their first derivatives for Co (a, b) and Fe (c, d) K-edges respectively, for all the samples. Our results suggest that the local chemical environment and the mean oxidation state are analogous for all samples, irrespective of the thermal treatment or Co/Fe ratio, indicated by the common features in the absorption edges: close the same pre-peaks position (associated to transitions Co/Fe 1s→ $e_g \uparrow$ and $e_g \downarrow$ transitions) and post edge features (corresponds to transition of electron from Co/Fe 1s → 4p orbital hybridized with different orbitals) could be shown in Figures 5a and 5c and the same behavior is found in the derivatives, Figures 5b and 5d, for the different compositions. The standard deviation for each one of the experimental points above the edge jump was less than 4%.

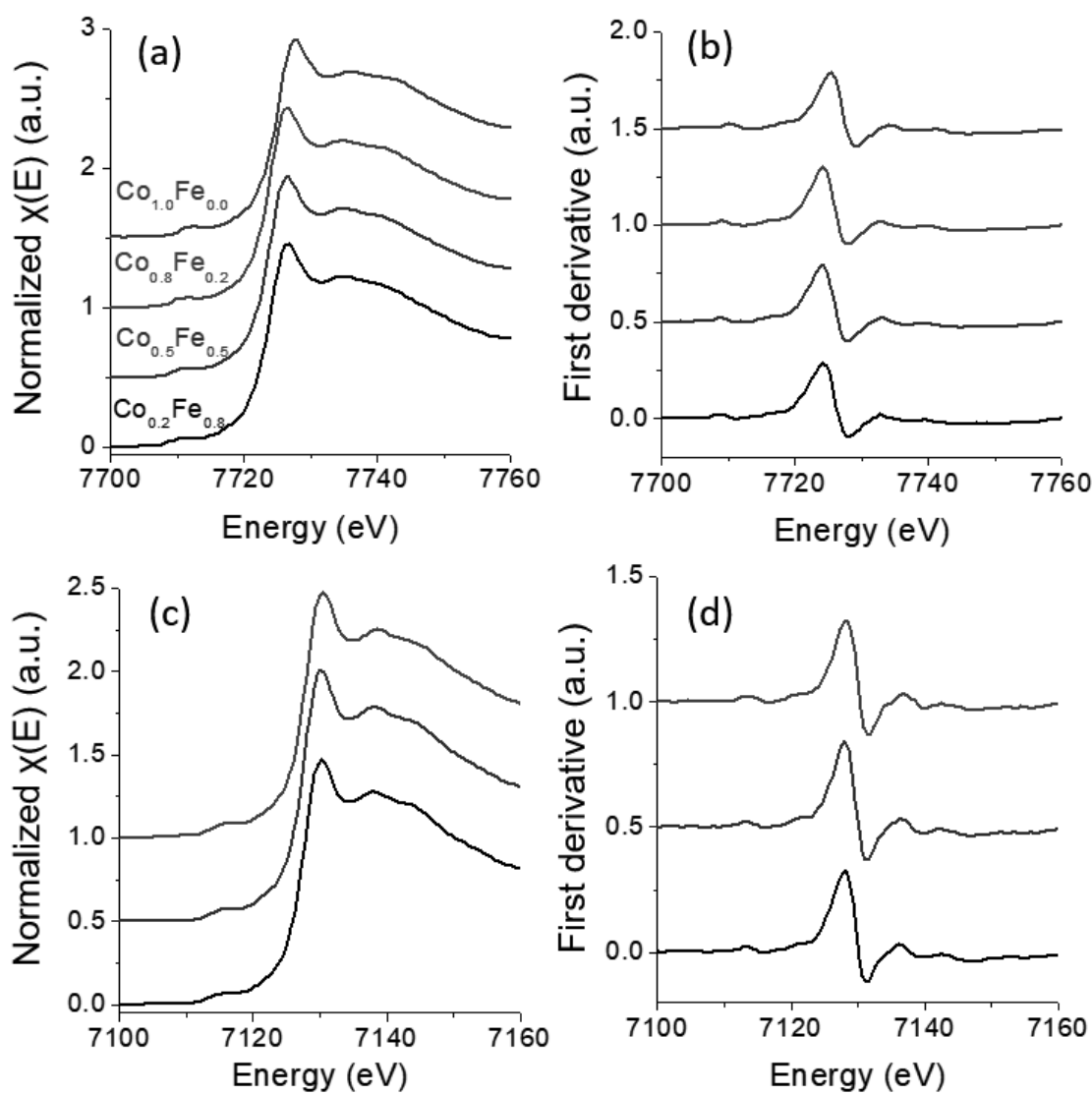

**Figure 5:** XANES average spectra and their first derivatives for Co (a,b) and Fe (c,d) K-edges: $La_{0.6}Sr_{0.4}Co_{0.8}Fe_{0.2}O_{3-\delta}$ in blue, $La_{0.6}Sr_{0.4}Co_{0.5}Fe_{0.5}O_{3-\delta}$ in wine and $La_{0.6}Sr_{0.4}Co_{0.2}Fe_{0.8}O_{3-\delta}$ in black lines.

The fact that K-edge spectra for Co and Fe showed similar features indicates that Co and Fe occupy a similar octahedral site in the structure, despite the change in the Co/Fe ratio. Pure $La_{0.6}Sr_{0.4}MO_3$ (M: Fe, Co) exhibits comparable characteristics [18,19] because the symmetry of the unit cell is determined by the La/Sr proportion [18].

The average oxidation state of Co and Fe ions were obtained by the integral methodology proposed by Capehart et al [20] where the energy axis from the flattened normalized data were shifted by placing the absorption edge at 0 eV. The corresponding values were calculated by calibrating the energy shift with respect to the origin between the sample and a metallic reference ($\delta E_S - \delta E_R$) at a selected area under the curve for different metallic oxides [21,22], as shown in Figure 6. A rebinding of data was required in order to minimize the integration error. The absorption coefficients of Co present a main jump between 7705 and 7725 eV. The calculated Co average oxidation states were close to 3.0 for all the Co/Fe relative ratios studied in this work. Meanwhile for Fe, the main absorption jump is in the range 7110-7130 eV, and the calculated Fe average oxidation states resulted close to 3.2 for the three compositions studied in this work. Table 3 summarizes the results of this analysis, including the values of the oxygen non-stoichiometry ($\delta$) of each composition determined imposing the neutrality condition.

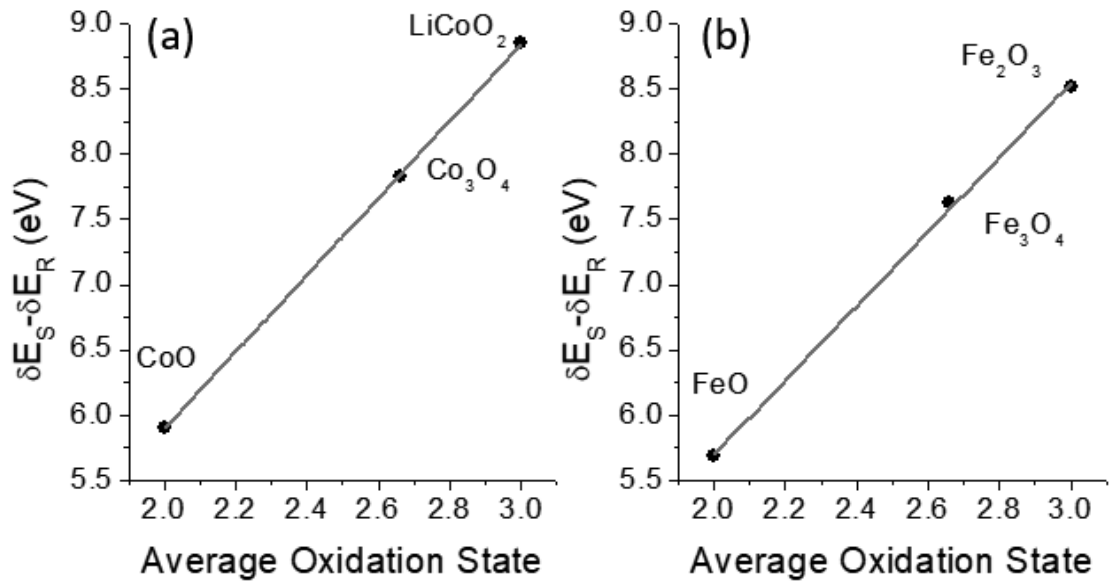

**Figure 6:** Calibration curves for XANES analysis determined using cobalt (a) and iron (b) oxides as references.

| Nominal composition | Co oxidation state | Fe oxidation state | $\delta$ |
|---|---|---|---|
| $La_{0.6}Sr_{0.4}Co_{0.8}Fe_{0.2}O_{3-\delta}$ | 2.96 (6) | 3.23 (9) | 0.19 (3) |
| $La_{0.6}Sr_{0.4}Co_{0.5}Fe_{0.5}O_{3-\delta}$ | 2.98 (6) | 3.22 (9) | 0.15 (4) |
| $La_{0.6}Sr_{0.4}Co_{0.2}Fe_{0.8}O_{3-\delta}$ | 3.01 (6) | 3.20 (9) | 0.12 (4) |

**Table 3:** Results of the XANES analysis for Co and Fe cations. The values of the oxygen non-stoichiometry ($\delta$) of each sample were determined imposing neutrality.

Figure 7 shows $\delta$ as a function of Fe content in the B site of the perovskite-type structure, showing a decreasing trend. The same behavior has been reported by other authors in similar materials, which can be attributed to the fact that Co on the B site has a smaller binding energy for oxygen compared to Fe [23,24,25]. This is probably related to the higher electrochemical performance found for $La_{0.6}Sr_{0.4}Co_{0.8}Fe_{0.2}O_{3-\delta}$ cathodes found in our previous work [4].

Remarkably, the values of oxygen non-stoichiometry determined in the present work for our nanostructured samples are much higher than those reported in the literature for conventional bulk materials [24,25,26,27,28]. This is probably due to the higher degree of disorder and structural defects of nanomaterials. This can also be related to the high performance of nanostructured mixed ionic-electronic conducting cathodes for the oxygen redox reaction, even though our factors can be also playing an importance role, such as the higher specific surface area and an enhanced grain boundary ionic diffusivity [1,2,3,4].

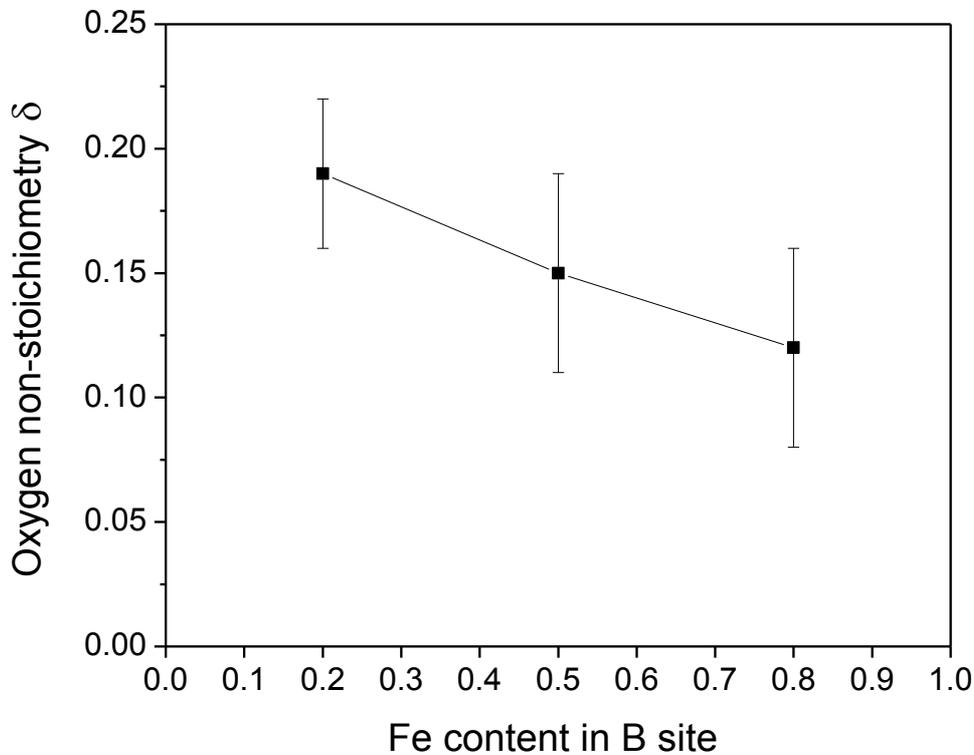

**Figure 7:** Oxygen non-stoichiometry of the LSCF samples studied in this work as a function of Fe content in the B site of the perovskite-type structure.

The weak Fe pre-edge peak observed in Figure 5 (c) is related to 1s–3d electric quadrupole transitions. This transition is forbidden by selection rules for metal ions with octahedral coordination, i.e. when there is an inversion center [29,30], but systems where the inversion symmetry is broken have dipole–quadrupole mixing [31]. The pre-edge features are sensitive to local symmetry [32], especially for $Fe^{+3}$ coordinated by six oxygen ions in an octahedral site [33]. Therefore, in the present work, Fe has been used as probe for octahedral environment symmetry. A decrease of the pre-edge contribution with an increase of iron concentration can be observed in Figure 8. The large pre-edge contribution for low Fe concentration indicates that centro-symmetry of the octahedral site is distorted. By replacing Co with Fe in the perovskite structure, the rhombohedral unit cell parameters changed unevenly and as a consequence, there is an enhancement in the point symmetry around the absorbing metal.

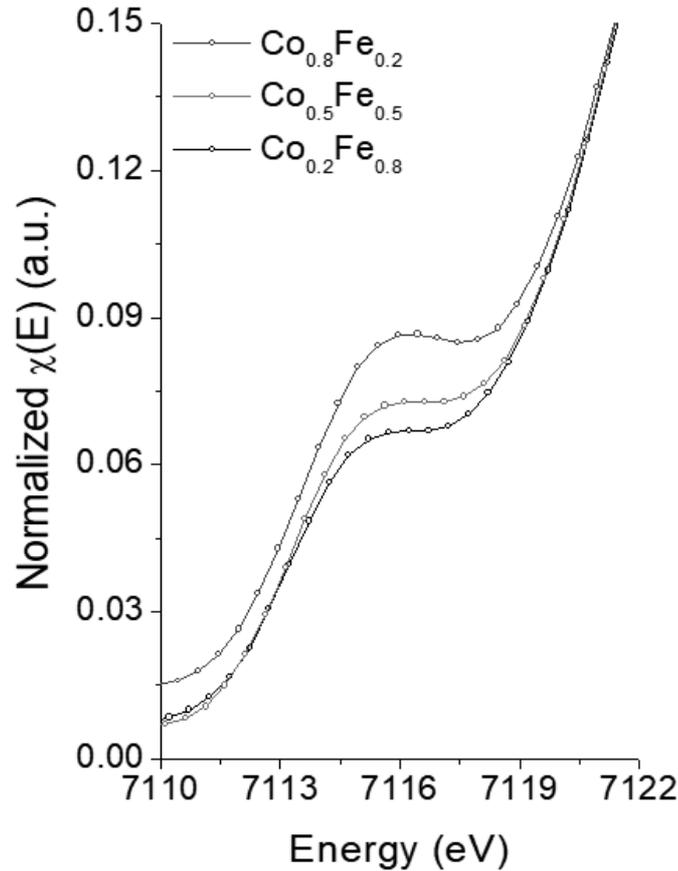

**Figure 8:** Fe K-edge XANES pre-edge spectra for different Fe concentrations.

*3.5. Mossbauer spectroscopy*

Figure 9 displays the Mössbauer spectra at room temperature in the large velocity range (10 mm/s). The richer Fe content samples (Fig. 9 (a) $La_{0.6}Sr_{0.4}Co_{0.2}Fe_{0.8}O_{3-\delta}$ and (b) $La_{0.6}Sr_{0.4}Co_{0.5}Fe_{0.5}O_{3-\delta}$) were fitted to two broad quadrupole split doublets (D1 and D2); meanwhile the less Fe-content sample was fitted to two doublet components (D1, D2) and a distribution of magnetic hyperfine field (S) (Fig. 9(c)). Hyperfine parameters are summarized in Table 4. Isomer shift (IS) values for D1 are characteristic of $Fe^{3+}$ and those for D2 are typical of $Fe^{4+}$ species.

It is noteworthy that the two richer Fe content samples are paramagnetic whilst the less Fe content sample has a magnetically ordered contribution at room temperature. Indeed, the spectrum of this sample reveals the reduction of the compound which is evident comparing the percent relative area of the spectra in Table 4. In this compound almost all $Fe^{4+}$ is reduced to $Fe^{3+}$, then the magnetic contribution may be due to the high strength of $Fe^{3+}$-$Fe^{3+}$ interaction which results in stronger superexchange interactions relative to those of $Fe^{4+}$-$Fe^{3+}$ [34]. Moreover, two different $Fe^{3+}$ contributions are distinguished from the $La_{0.6}Sr_{0.4}Co_{0.8}Fe_{0.2}O_{3-\delta}$ spectrum, denoting two types of $Fe^{3+}$ sites. The sextet has an IS characteristic of $Fe^{3+}$ in octahedral coordination while that of the D1 doublet is characteristic of $Fe^{3+}$ in coordination lower than octahedral (usually observed in oxygen-deficient perovskites [35]). This information is in tune with the change in symmetry suggested by XANES results. Since SXPD data confirmed that all the samples are single-phased, the presence of these two types of $Fe^{3+}$ sites can be understood in view of the fact that Mössbauer spectroscopy is sensitive to the local structure while SXPD studies the long-range atomic order.

The fact that the IS of the $Fe^{3+}$ doublet varies with Fe content also points out to a possible different degree of oxygen deficiency in the samples [36]. When the compound reduces, oxygen deficiency should increase. Consequently, $La_{0.6}Sr_{0.4}Co_{0.2}Fe_{0.8}O_{3-\delta}$ and $La_{0.6}Sr_{0.4}Co_{0.5}Fe_{0.5}O_{3-\delta}$ samples which show paramagnetic spectra would suggest oxygen-rich Fe environments relative to that of the $La_{0.6}Sr_{0.4}Co_{0.8}Fe_{0.2}O_{3-\delta}$ sample. From Table 4 it is also seen that the estimated line width values for the $Fe^{3+}$ doublet component increase with decreasing Fe concentration. Once again, this may result from increasing disorder due to an increasing oxygen-vacancy concentration.

Average iron oxidation states were calculated from relative areas of the contributions in the spectra, obtaining $Fe^{3.3+}$ for the $La_{0.6}Sr_{0.4}Co_{0.2}Fe_{0.8}O_{3-\delta}$ and the $La_{0.6}Sr_{0.4}Co_{0.5}Fe_{0.5}O_{3-\delta}$ sample and $Fe^{3.04+}$ for the $La_{0.6}Sr_{0.4}Co_{0.8}Fe_{0.2}O_{3-\delta}$ sample.

| Nominal composition | | Oxidation state | W [mm/s] | IS [mm/s] | QS [mms] | $B_{hf}$ [T] | $2\varepsilon$ [mm/s] | Relative area [%] |
|---|---|---|---|---|---|---|---|---|
| $La_{0.6}Sr_{0.4}Co_{0.2}Fe_{0.8}O_{3-\delta}$ | D1 | $Fe^{3+}$ | 0.31 | 0.29 | 0.28 | - | - | 69 |
| | D2 | $Fe^{4+}$ | 0.29 | 0.06 | 0.26 | - | - | 31 |
| $La_{0.6}Sr_{0.4}Co_{0.5}Fe_{0.5}O_{3-\delta}$ | D1 | $Fe^{3+}$ | 0.43 | 0.27 | 0.30 | - | - | 70 |
| | D2 | $Fe^{4+}$ | 0.27 | 0.12 | 0.15 | - | - | 30 |
| $La_{0.6}Sr_{0.4}Co_{0.8}Fe_{0.2}O_{3-\delta}$ | D1 | $Fe^{3+}$ | 0.65 | 0.22 | 0.25 | - | - | 32 |
| | D2 | $Fe^{4+}$ | 0.25 | 0.12 | 0.16 | - | - | 4 |
| | S | $Fe^{3+}$ | - | 0.37 | - | 40.0 | -0.11 | 64 |

**Table 4:** Hyperfine parameters at room temperature. Paramagnetic doublets (D1 and D2) - W: line widths, IS: isomer shift, QS: quadrupole splitting. Magnetically split sextet (S) - $B_{hf}$: average magnetic hyperfine field, $2\varepsilon$: quadrupole shift.

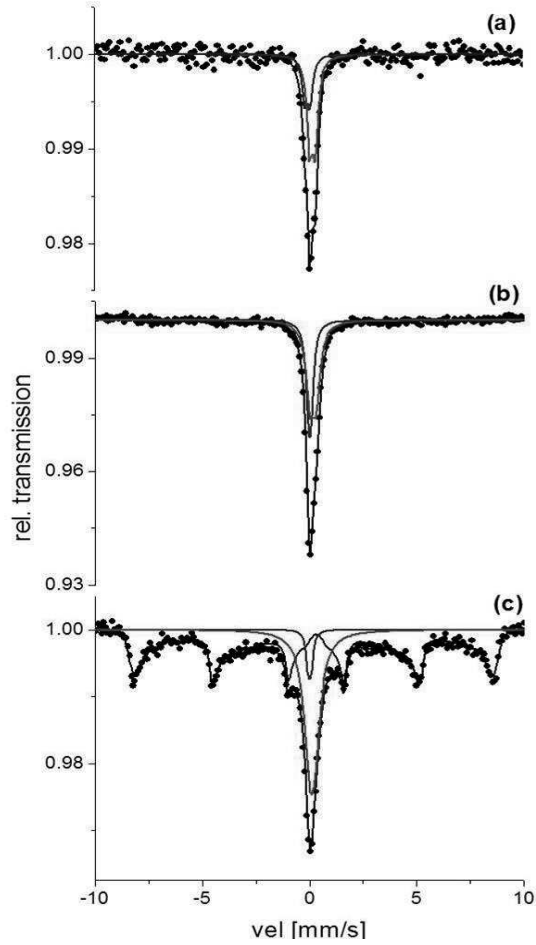

**Figure 9:** Fitted Mössbauer spectra corresponding to (a) $La_{0.6}Sr_{0.4}Co_{0.2}Fe_{0.8}O_{3-\delta}$, (b) $La_{0.6}Sr_{0.4}Co_{0.5}Fe_{0.5}O_{3-\delta}$ and (c) $La_{0.6}Sr_{0.4}Co_{0.8}Fe_{0.2}O_{3-\delta}$, respectively.

## 4. Conclusions

The present study shows that single-phased LSCF nanorods with rhombohedral perovskite-type phase can be obtained by the proposed pore-wetting technique. An increase of the lattice parameters was observed with increasing Fe content, as expected according to the data reported in the literature for the pure Co and Fe samples. Nanometric crystallites of around 25 nm mean diameter were obtained. It is worth to point out that, even though several entries can be found in crystallographic databases for LSCF materials with similar compositions, only a few were measure at room-temperature and none of them correspond to nanocrystalline samples.

The resulting material showed a rod-like submicrometric structure, which has been shown to be optimum to develop cathodes for IT-SOFCs. The particles that form those rods are of around 140 nm of diameter in average, so these particles are formed by several nanocrystals.

The results from XANES spectroscopy showed that the Co and Fe average oxidation states and chemical environment are similar for all samples. The fact that both K-edge spectra are similar also indicates that Co and Fe occupy the same octahedral site. The changes in the pre-edge of the spectra are consistent with the substitution of Co by Fe. The average oxidation states resulted close to 3 and 3.2 for Co and Fe, respectively, for all the compositions studied in this work.

From the results of XANES analysis, the oxygen non-stoichiometry $\delta$ of the samples was determined, finding that it decreases with increasing Fe content, as expected according to the higher Fe-O bond energy comparted to that of Co-O one. The values of $\delta$ found in this work are much higher than those reported in the literature for bulk materials, probably due to the high degree of defects of nanomaterials. This can be related to the excellent electrochemical performance of nanorod-like cathodes [4].

The average oxidation states of Fe obtained from Mössbauer spectroscopy resulted in agreement with those obtained by XANES. By this technique, we also obtained evidence of two types of $Fe^{3+}$ ions, one in octahedral coordination and the other with lower coordination, a typical characteristic of oxygen deficient samples. Regarding the magnetic interactions, a paramagnetic to a partially magnetic ordered behavior is observed from rich to poor Fe content samples.

**Acknowledgements**

The present work was partially supported by the Brazilian Synchrotron Light Laboratory (LNLS, Brazil, proposals XPD 10152 and XAFS1 9886), Agencia Nacional de Promoción Científica y Tecnológica (Argentina, PICT 2015 No. 3411), CONICET (Argentina, PIP 00362) and CAPES-MinCyT bilateral cooperation.